\documentclass[letter,twocolumn]{jpsj2}

\title{Independent Component Analysis of Spatiotemporal Chaos}

\author{
  Hirokazu \textsc{Asano}$^{1}$ and Hiroya
  \textsc{Nakao}$^{2}$\thanks{Corresponding author: nakao@ton.scphys.kyoto-u.ac.jp}
}

\inst{
  $^{1}$Department of Applied Mathematics and Physics,
  Graduate School of Informatics, Kyoto University, Kyoto 606-8501, Japan\\
  $^{2}$Department of Physics, Graduate School of Science, Kyoto
  University, Kyoto 606-8502, Japan
}

\abst{
  Two types of spatiotemporal chaos exhibited by ensembles of coupled
   nonlinear oscillators are analyzed using independent component
  analysis (ICA).  For diffusively coupled complex Ginzburg-Landau
  oscillators that exhibit smooth amplitude patterns, ICA extracts
  localized one-humped basis vectors that reflect the characteristic
  hole structures of the system, and for nonlocally coupled complex
  Ginzburg-Landau oscillators with fractal amplitude patterns, ICA
  extracts localized basis vectors with characteristic gap structures.
  Statistics of the decomposed signals also provide insight into the
  complex dynamics of the spatiotemporal chaos.  }

\recdate{\today}

\kword{coupled oscillators, spatiotemporal chaos, signal processing,
  feature extraction}

\begin{document}

\maketitle

Spatiotemporal chaos (STC) arising from interaction between autonomous
elements is ubiquitous in nonequilibrium dissipative systems such as
fluid flows and chemical reactions~\cite{LCGL1,LCGL2,LCGL3,LCGL4}.
It is generally difficult to reduce the complex behavior of strongly
coupled systems to the individual dynamics of the component elements,
so we must employ collective modes of the system for description.  For
example, we employ various types of bases in the description of
fluids, e.g. Fourier basis, wavelet basis, or eigenfunctions of
linearized evolution equations, which represent flow modes such as
waves, vortices, and
convections~\cite{LCGL1,LCGL2,LCGL3,LCGL4,Flow1,Flow2,Mallat}. In this
case, the observer needs to subjectively choose which basis to use
depending on the situation.

On the other hand, there also exist methods of constructing basis
functions objectively from statistical properties of the system by
some criterion, without explicitly fixing them. A representative
method is the principal component analysis (PCA) or the
Karhunen-Lo\'eve
expansion~\cite{Flow1,Flow2,Mallat,ICA1,ICA2,ICA3,ICA4,ICA5,PCA1,PCA2,PCA3,PCA4,PCA5},
which is a standard method in multivariate analysis.
PCA has already been employed in analyzing STC, for example, to
truncate its evolution equation or to estimate its degree of
freedom~\cite{PCA1,PCA2,PCA3,PCA4,PCA5}.
However, as we show below, when PCA is applied to STC, spatially
delocalized bases are typically extracted, which are not appropriate
for describing local field structures. Particularly, when the system
has translational symmetry, it can be proven that PCA extracts the
Fourier basis itself.
Therefore, some statistical method that can objectively and compactly
capture the complex field structures is desirable.

In this letter, we apply independent component analysis (ICA) to
STC. ICA is a recently developed method of statistical signal
processing, which attempts to decompose observed mixed signals into
maximally independent signals~\cite{ICA1,ICA2,ICA3,ICA4,ICA5}.
It is known that the ICA basis gives an information-theoretically
efficient representation of multivariate
signals~\cite{ICA1,ICA2,ICA3,ICA4,ICA5}.
We analyze STC exhibited by two types of coupled nonlinear
oscillators, namely, locally (diffusively) coupled complex
Ginzburg-Landau oscillators (LCGL)~\cite{LCGL1,LCGL2,LCGL3,LCGL4}, and
nonlocally coupled complex Ginzburg-Landau oscillators
(NCGL)~\cite{NCGL1,NCGL2,NCGL3,NCGL4,NCGL5}, and assess the utility of ICA in analyzing such
STC.

\begin{figure}[htbp]
  \begin{center}
    \includegraphics[width=0.95\hsize]{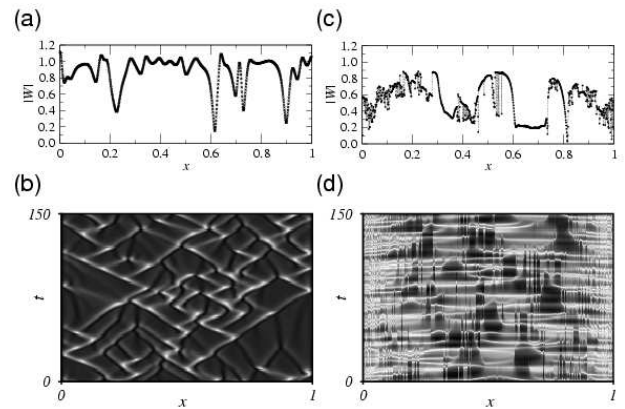}
  \end{center}
  \caption{(a) Snapshot and (b) evolution of the amplitude patterns
    exhibited by LCGL. (c) Snapshot and (d) evolution of the amplitude
    patterns exhibited by NCGL.}
  \label{fig_snaps}
\end{figure}

We consider spatially one-dimensional systems of unit length on
$[0,1]$.
The LCGL is a system of diffusively coupled CGL oscillators, which
obeys
\begin{equation}
  \frac{\partial W(x,t)}{\partial t} =
  W - (1 + i c_2) |W|^2 W + D (1 + i c_1) \frac{\partial^2 W}{\partial x^2},
  \label{Eq:LCGL}
\end{equation}
where $x$ represents the position, $t$ the time, and $W$ the complex
amplitude of the CGL oscillator. This is a universal equation derived
from general evolution equations for self-oscillatory media by the
center-manifold reduction near the supercritical Hopf bifurcation
point~\cite{LCGL1,LCGL2,LCGL3,LCGL4}. Here, $c_2$ represents the
angular velocity of a single oscillator, $c_1$ the phase shift of
diffusive coupling, and $D$ the diffusion coefficient. We fix these
parameters at $c_1 = 1.0$, $c_2 = -1.5$, and $D=5 \times 10^{-5}$.
We assume Neumann boundary conditions $\partial W / \partial x|_{x=0}
= \partial W / \partial x|_{x=1} = 0$, so that the system is not
translationally symmetric.
When $c_1$ and $c_2$ satisfy the Benjamin-Feir condition $1+c_1c_2 <
0$, the system exhibits STC.
Figure~\ref{fig_snaps}(a) shows a typical snapshot of the amplitude
pattern $|W(x,t)|$ and Fig.~\ref{fig_snaps}(b) its temporal
evolution. In this parameter region, the amplitude pattern is composed
of characteristic hole structures that move around the system randomly
(defect turbulence~\cite{LCGL1,LCGL2,LCGL3,LCGL4,Hole1,Hole2,Hole3,Hole4}).

\begin{figure}[htbp]
  \begin{center}
    \includegraphics[width=0.5\hsize]{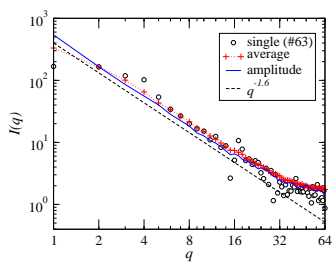}
  \end{center}
  \caption{Power spectra of the original amplitude pattern and the ICA
    basis vectors of the NCGL. Circles($\circ$) represent the spectrum
    of a single ICA basis vector, crosses($+$) the averaged spectrum
    over all ICA basis vectors, and the solid line the spectrum
    directly calculated from the amplitude pattern. The broken line is
    the power-law function $q^{-1.6}$. Each curve is vertically
    shifted arbitrarily for comparison.}
  \label{fig_ncgl_power}
\end{figure}

On the other hand, the NCGL is a system of CGL oscillators coupled
through a nonlocal mean field weighted by a kernel that decreases
exponentially with distance~\cite{NCGL1,NCGL2,NCGL3,NCGL4,NCGL5}. It
obeys
\begin{eqnarray}
  \frac{\partial W(x,t)}{\partial t} &=& W - (1 + i c_2) |W|^2 W + K(1 +
  ic_1)\cr
  &&\times \left\{ \frac{\gamma}{2} \int_{0}^{1} e^{- \gamma |x'-x|}
    W(x',t) dx' - W \right\},\;\;\;\;\;
  \label{Eq:NCGL}
\end{eqnarray}
where $c_2$ represents the angular velocity, $c_1$ the phase shift of
nonlocal coupling, $K$ the coupling strength, and $\gamma^{-1}$ the
coupling distance. We fix the parameters at $c_1 = -2.0$, $c_2 = 2.0$,
$\gamma^{-1} = 0.125$, and $K=0.85$.
We do not assume periodic boundary conditions, and restrict the
integration interval to $[0,1]$ (though the kernel is formally
normalized in the infinite domain).
When the Benjamin-Feir condition $1+c_1c_2 < 0$ holds, this system
also exhibits STC. However, this STC is considerably different from
that of the LCGL.
Figure~\ref{fig_snaps}(c) shows a typical snapshot of the amplitude
pattern $|W(x,t)|$ and Fig.~\ref{fig_snaps}(d) its temporal
evolution. The amplitude pattern is not continuous but consists of
relatively coherent regions and disordered regions with discontinuous
gaps. Its power spectrum $I(q)$ exhibits a power-law dependence on the
wave number $q$ as shown in
Fig.~\ref{fig_ncgl_power}~\cite{NCGL1,NCGL2,NCGL3,NCGL4,NCGL5}. In
this parameter region, such a fractal disordered amplitude pattern
evolves steadily.

We focus on the amplitude pattern $|W(x,t)|$ of these two systems
hereafter.
We discretize a certain domain of space into $n$ segments of length
$\Delta x$ and define $u_i(t) = |W(i \Delta x, t)|$.
The amplitude pattern is now represented by an $n$-dimensional vector
$\textbf{u}(t) = ( u_1(t), \cdots, u_n(t) )^T$, and its temporal
sequence is observed.
We consider the observed signal and its transformations (denoted by
lower-case letters, e.g., ${\bf u}(t)$ or ${\bf z}(t)$), to be the
realizations of the underlying stochastic variables (denoted by
capital letters, e.g., ${\bf U}$ or ${\bf Z}$).
In the following, we subtract the mean value $E[{\textbf{U}}]$ from
$\textbf{u}(t)$ and regard $\textbf{z}(t) = (z_1(t), \cdots, z_n(t))^T
= \textbf{u}(t) - E[{\textbf{U}}]$ as the observed signal. Here,
$E[\cdots]$ denotes the expectation, which is substituted by the
long-time average of the observed signal in actual numerical
calculations.
We decompose $\textbf{z}(t)$ using an $n$-dimensional basis
$\{\textbf{a}_j \}$ ($j = 1 \cdots n$) as
\begin{equation}
  \textbf{z}(t) = \sum^{n}_{j=1} s_j(t) \textbf{a}_j \label{expansion2},
\end{equation}
where $\textbf{s}(t) = (s_1(t), \cdots, s_n(t))^T$ represents the
decomposed signals.
If we use the sinusoidal basis here, this gives a Fourier
decomposition.  In the following, we use PCA and ICA bases.

PCA extracts orthogonal basis vectors $\{\textbf{a}_j \}$ on which the
stochastic variable $\textbf{Z}$ distributed in the $n$-dimensional
space has the maximum projection (principal components) in order of
importance. PCA is achieved by choosing the eigenvectors of the
covariance matrix $\boldsymbol{\Sigma}$ of $\textbf{Z}$ as the
basis~\cite{Flow1,Flow2,Mallat,ICA1,ICA2,ICA3,ICA4,ICA5}.
Namely, each $\{\textbf{a}_j \}$ satisfies an eigenequation $
\boldsymbol{\Sigma} \textbf{a}_j = \lambda_j \textbf{a}_j $, where $
\boldsymbol{\Sigma} = E[\textbf{Z} \textbf{Z}^T] = \left( E[Z_i Z_j]
\right) $.
Since $\boldsymbol{\Sigma}$ is a real symmetric matrix, $\lambda_j
\geq 0$ for all $j$. Each $\lambda_j$ represents the relative ratio of
the $j$-th component $\textbf{a}_j$ contained in the observed data.
The coefficients $s_i(t)$ and $s_j(t)$ are uncorrelated when $i \neq
j$, but generally not independent.

On the other hand, ICA attempts to extract mutually independent
signals $\textbf{s}(t)$ from the observed signals $\textbf{z}(t)$,
which are assumed to be a linear mixture of $\textbf{s}(t)$ by some
unknown constant $n \times n$ matrix $\textbf{A}$, i.e.,
$\textbf{z}(t) = \textbf{A} \textbf{s}(t)$.
We assume $\textbf{s}(t)$ to be the outcomes of a stochastic variable
$\textbf{S}$, whose components are mutually independent.
It is known that~\cite{ICA1,ICA2,ICA3,ICA4,ICA5} if the number of
Gaussian-distributed components of $\textbf{S}$ is at most one (which
we assume hereafter), a decomposing matrix $\textbf{W}$ exists such
that the decomposed signal $ \textbf{y}(t) = (y_1(t), \cdots,
y_n(t))^T = \textbf{W}\textbf{z}(t) $ coincides with $\textbf{s}(t)$,
except for inevitable ambiguities in the scale and order of the
components.
To find such a decomposing matrix $\textbf{W}$, we minimize the mutual
information $I(\textbf{Y})$ between the probability density function
(PDF) $P(\textbf{y})$ of the stochastic variable $\textbf{Y} = (Y_1,
\cdots, Y_n)^{T}$ and its marginalized PDFs $P_{i}(y_i) = \int
P(\textbf{y}) dy_1 \cdots dy_{i-1} dy_{i+1} \cdots dy_n$,
\begin{align}
  I(\textbf{Y}) = \int P(\textbf{y}) \log \left( \frac{P(\textbf{y})}
    {P_{1}(y_1) \cdots P_{n}(y_n)} \right) d\textbf{y},
  \label{mutualinformation}
\end{align}
so that the decomposed signals become maximally independent.
The matrix $\textbf{W}$ that minimizes $I(\textbf{Y})$ can be obtained
using Amari's natural gradient method~\cite{ICA1,ICA2,ICA3,ICA4,ICA5}
as
\begin{align}
  \textbf{W} \leftarrow \textbf{W} - \eta \frac{\partial
    I(\textbf{Y})}{\partial \textbf{W}}\textbf{W}^T \textbf{W} =
  \textbf{W} + \eta \left(
    \textbf{I}-E[\boldsymbol{\varphi}(\textbf{Y}) \textbf{Y}^T]
  \right) \textbf{W},
  \label{koubai2}
\end{align}
where $\eta$ is a learning rate, and $\boldsymbol{\varphi}(\textbf{Y})
= (\varphi_1(Y_1), \cdots, \varphi_n(Y_n))^{T}$ is a nonlinear vector
function of $\textbf{Y}$.
By formally differentiating eq.~(\ref{mutualinformation}), each
component of $\boldsymbol{\varphi}(\textbf{Y})$ is calculated as $
\varphi_{i}(Y_i) = - \partial \log P_{i}(Y_i) / \partial Y_i $, which
necessitates knowledge of the unknown functions $P_{i}(Y_i)$.
However, it is known that~\cite{ICA1,ICA2,ICA3,ICA4,ICA5} the signals
can be decomposed by assuming simple functional forms for
$\varphi_{i}(Y_i)$ instead of using the true PDF of $Y_i$, and various
algorithms have been devised.
In the following, we use the well-known extended infomax
algorithm~\cite{ICA1,ICA2,ICA3,ICA4,ICA5} given by
\begin{align}
  &\varphi_{i}(Y_i) = k_i\tanh(Y_i) + Y_i,\cr
  &k_i = \mbox{sign} \left\{ E[1-\tanh ^2(Y_i)] E[Y_i^2] -
    E[\tanh(Y_i) Y_i] \right\},
  \;\;\;\;\;
\end{align}
where $k_i$ takes either $+1$ or $-1$ depending on the sign of the
expression inside the curly brackets.
From the estimated decomposing matrix $\textbf{W}$, ICA basis vectors
are obtained as $(\textbf{a}_1,\cdots,\textbf{a}_n) = \textbf{W}^{-1}
$. When $i \neq j$, the coefficients $s_i(t)$ and $s_j(t)$ are
mutually independent, but $\textbf{a}_i$ and $\textbf{a}_j$ are
generally not orthogonal.

In actual numerical calculations, the LCGL and NCGL are numerically
simulated using $N=2^{10} - 2^{11}$ spatial grid points.  We
discretize the central part of the system $0.375 \leq x < 0.625$ using
$n=128$ points ($\Delta x = 1/512$) and observe the amplitude pattern
every second for $20000$ seconds. The decomposing matrix $\textbf{W}$
is initially set to an identity matrix and updated $5000 - 10000$
times using a learning rate $\eta$ between $0.05 - 0.3$.
Due to the ambiguity of ICA, the extracted vectors depend on the
observed data and on the initial condition for $\textbf{W}$, even if
the parameters of the systems are the same. Also, we can generally
find only local minima in this type of high-dimensional (here
$128^2$-dimensional) optimization problem.
However, in our numerical calculations, we always obtained
qualitatively the same results for various data sets, including those
obtained using cyclic boundary conditions.

\begin{figure}[htbp]
  \begin{center}
    \includegraphics[width=0.95\hsize]{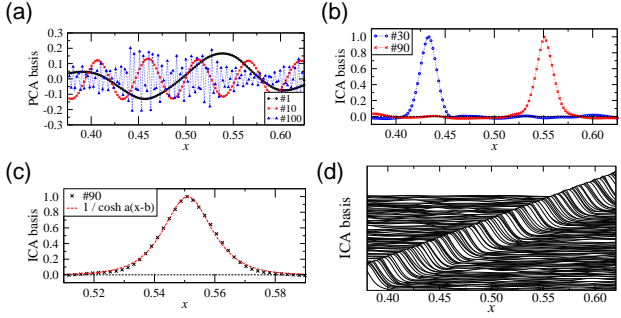}
  \end{center}
  \caption{(a) PCA basis vectors No. $1$, $10$, and $100$.  (b) ICA
    basis vectors No. $30$ and $90$. (c) Magnification of ICA basis
    No. $90$. The dotted line is a fit using the function $1 / \cosh
    [a(x-b)]$, where the parameters are estimated to be $a = 141$ and
    $b = 0.551$. (d) All ICA basis vectors. The graph is plotted
    stereoscopically by gradually shifting the baseline of each basis.}
  \label{fig_lcgl_basis}
\end{figure}

 Let us analyze the LCGL first. Figure~\ref{fig_lcgl_basis}(a) shows
three typical basis vectors among $128$ vectors obtained by PCA. These
vectors, as well as the other vectors not displayed here, are all
roughly sinusoidal and spatially delocalized.
Thus, the PCA basis does not capture the local hole structures of the
LCGL very well.
Figure~\ref{fig_lcgl_basis}(b) shows two typical basis vectors among
$128$ vectors obtained by ICA. Reflecting the characteristic hole
structures of the system, spatially localized one-humped vectors are
extracted. Each vector can be fitted well by a function given by $1 /
\cosh [a(x-b)]$, as shown in Fig.~\ref{fig_lcgl_basis}(c).
There is no a priori reason for it to satisfy the LCGL as a special
solution~\cite{Hole1,Hole2,Hole3,Hole4}, but its width is roughly the
same as that of the actual hole structures and scales properly with
the diffusion constant $D$.
All other basis vectors have almost the same shape, but their
locations differ from each other, so that the entire observed domain
is completely covered, as shown in Fig.~\ref{fig_lcgl_basis}(d).
Here, the obtained basis vectors are sequentially aligned because we
used an identity matrix as the initial $\textbf{W}$. If we use a
random matrix as the initial $\textbf{W}$, we obtain similar basis
vectors but their order and vertical orientation become random.

\begin{figure}[htbp]
  \begin{center}
    \includegraphics[width=0.95\hsize]{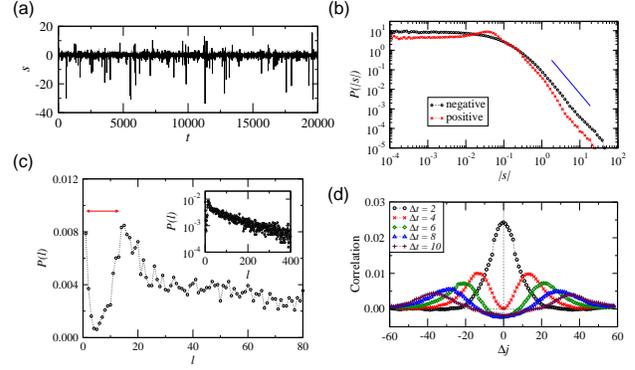}
  \end{center}
  \caption{(a) Temporal sequence of the coefficient $s$ of the ICA
    basis vector No. $90$. (b) PDFs $P(|s|)$ of absolute coefficient
    $|s|$ for $s>0$ and $s<0$. (c) PDF $P(l)$ of interval $l$ between
    two successive generation of holes (short time-scale region). The
    inset shows $P(l)$ at a larger scale. (d) Correlation $C(\Delta j,
    \Delta t)$ between two decomposed signals.}
  \label{fig_lcgl_tseq}
\end{figure}

Let us turn our attention to the ICA expansion coefficients
(decomposed signals). Since the ICA basis reflects localized hole
structures, when the coefficient $s_j$ of some basis vector
$\textbf{a}_j$ takes a large negative value, we can judge that there
exists a hole at position $j$. The temporal evolution of a coefficient
$s$ corresponding to a typical ICA basis vector (No. $90$) is shown in
Fig.~\ref{fig_lcgl_tseq}(a). Corresponding to the appearance of holes,
$s$ occasionally takes a large negative value.
Figure~\ref{fig_lcgl_tseq}(b) shows PDFs $P(|s|)$ of the absolute
value $|s|$ averaged over all signals in logarithmic scales for $s>0$
and $s<0$. Both distributions are super-Gaussian with much heavier
tails than the normal distribution. Reflecting the concave asymmetric
shape of the vector, $P(|s|)$ for positive $s$ and negative $s$ are
different. The tail part of $P(|s|)$ exhibits power-law behavior,
implying the existence of some self-similarity in the formation of
holes.

To characterize the temporal structure of $s$, we investigate the PDF
of time interval $l$ between two successive events at which $s$
becomes smaller than some negative threshold $h$. It represents the
time interval between two holes, where $h$ determines how large an
event must be to be counted as a hole. Figure~\ref{fig_lcgl_tseq}(c)
displays the PDF $P(l)$ of $l$ for $h=-3$ averaged over all signals.
The PDF dips considerably when $0 < l < 20$ (shown by an arrow), which
indicates the existence of refractory periods. After the disappearance
of the first hole, the next hole is less likely to appear at the same
position for awhile. The PDF decreases exponentially at large values
of $l$ as shown in the inset, indicating a Poissonian random
appearance of holes at large time scales.

\begin{figure}[htbp]
  \begin{center}
    \includegraphics[width=0.95\hsize]{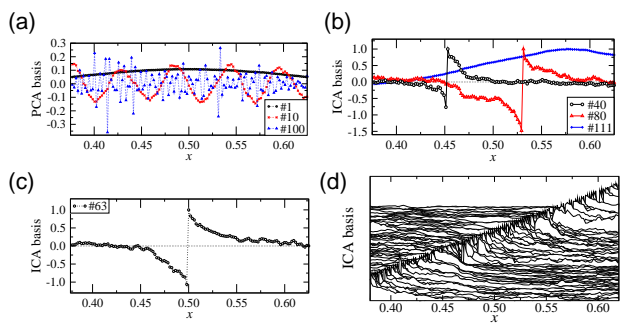}
  \end{center}
  \caption{(a) PCA basis vectors No. $1$, $10$, and $100$. (b) ICA
    basis vectors No. $40$, $80$, and $111$. (c) Magnification of ICA
    basis vector No. $63$. (d) All ICA basis vectors, drawn
    stereoscopically by shifting the baselines. A few long-wavelength
    vectors are omitted.}
  \label{fig_ncgl_basis}
\end{figure}

Figure~\ref{fig_lcgl_tseq}(d) shows correlation functions $C(\Delta j,
\Delta t) = \overline{ E[ s_{j}(t) s_{j+\Delta j}(t+\Delta t) ] }$
between two signals $s_{j}$ and $s_{j+\Delta j}$ whose index numbers
are separated by $\Delta j$ for several values of the temporal
difference $\Delta t$.  The curves are averaged over $j$ and $t$, and
smoothed by averaging over $5$ neighboring data points on both sides.
Since the vectors are spatially aligned sequentially, $\Delta j$
corresponds to a spatial distance between two basis vectors.  The
correlation gradually spreads from neighboring vectors to distant
vectors, which clearly captures the dynamics and decoherence of hole
structures. Note that the ICA algorithm used here does not take into
account the temporal structure of the signals, hence the decomposed
signals can be correlated when $\Delta t \neq 0$.

Let us now analyze the NCGL. Examples of PCA basis vectors are shown
in Fig.~\ref{fig_ncgl_basis}(a). All PCA vectors are delocalized and
sinusoidal in this case again. Figures~\ref{fig_ncgl_basis}(b) and
\ref{fig_ncgl_basis}(c) show ICA basis vectors. As shown in
Fig.~\ref{fig_ncgl_basis}(c), most vectors typically have a localized
structure with a gap and two tails, reflecting the gaps of the
amplitude pattern of the NCGL. We also found a few (at most $5$
percent of the total number) long-wavelength delocalized vectors such
as No. $111$ shown in Fig.~\ref{fig_ncgl_basis}(b). In the following
analysis, we omit such exceptional vectors.
The location of each vector is different, and the entire observed
domain is covered, as shown in Fig.~\ref{fig_ncgl_basis}(d).  These
ICA basis vectors not only reflect the gap structures of the NCGL but
also reflect its singularity.  As shown in Fig.~\ref{fig_ncgl_power},
the power spectrum of the ICA basis vector exhibits a power law
similar to that of the original amplitude pattern. Although the
statistics are poor, the power-law behavior can readily be seen for a
single basis. The average spectrum over all vectors agrees well with
the spectrum of the original amplitude pattern. In these respects, the
ICA basis captures the given spatial patterns more faithfully than the
PCA basis, which does not possess such properties.

\begin{figure}[htbp]
  \begin{center}
    \includegraphics[width=0.95\hsize]{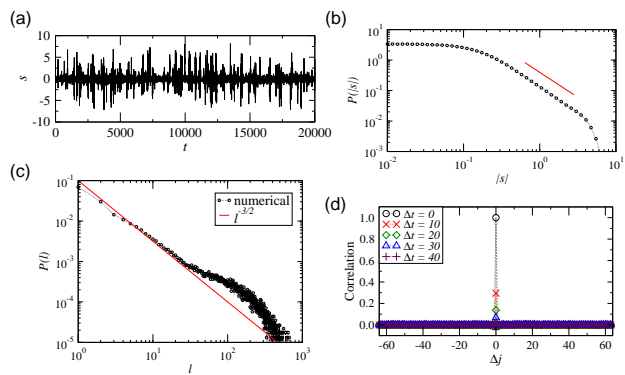}    
  \end{center}
  \caption{ (a) Time sequence of the coefficient $s$ of the ICA basis
    vector No. $63$. (b) PDFs $P(|s|)$ of the coefficient $|s|$. (c)
    PDFs $P(l)$ of the laminar length $l$. The straight line is a
    power-law $l^{-3/2}$. (d) Correlation $C(\Delta j, \Delta t)$
    between two decomposed signals.}
  \label{fig_ncgl_tseq}
\end{figure}

Since each ICA basis vector has a gap structure, its coefficient
quantifies a gap in the amplitude pattern. It was shown in previous
studies~\cite{NCGL1,NCGL2,NCGL3,NCGL4,NCGL5} that the difference in
amplitude between two nearby oscillators exhibits noisy on-off
intermittency~\cite{OnOff1,OnOff2}.  Thus, the ICA coefficient is also
expected to exhibit noisy on-off intermittency.
Figure~\ref{fig_ncgl_tseq}(a) shows the temporal evolution of a
typical ICA coefficient $s$ (No. $63$). Large gap structures are
generated intermittently.
The PDF $P(|s|)$ of the absolute coefficient $|s|$ averaged over all
signals is shown in Fig.~\ref{fig_ncgl_tseq}(b). In this case,
$P(|s|)$ for both $s>0$ and $s<0$ is symmetric, since the shape of
each basis vector is statistically symmetric. The PDF is
super-Gaussian with a flat region for small $|s|$, a power-law region
for medium $|s|$, and a sharp cutoff due to nonlinearity at large
$|s|$. This shape is characteristic of the noisy on-off
intermittency~\cite{NCGL1,NCGL2,NCGL3,NCGL4,NCGL5}.
Figure~\ref{fig_ncgl_tseq}(c) shows the PDF of the laminar length
interval $l$ during which $s$ is smaller than some threshold $h$ (here
$h=1.5$). It represents the duration between two successive
appearances of gap structures.
It is known that the laminar length PDF of noisy on-off intermittency
is determined by the first passage time PDF of a Wiener process, which
has a power-law region $l^{-3/2}$ and an exponential
cutoff~\cite{OnOff1,OnOff2}. This is confirmed in
Fig.~\ref{fig_ncgl_tseq}(c) for the ICA coefficient.
Finally, Fig.~\ref{fig_ncgl_tseq}(d) shows the correlation $C(\Delta
j, \Delta t)$ between signals for several values of $\Delta t$. The
correlation does not spread at all in this case, indicating that the
ICA basis vectors persist to be independent of each other for a long
duration.

Summarizing, we applied ICA to STC in two types of coupled oscillators
and extracted localized basis vectors that accurately represent the
local structures of those systems, which demonstrates the utility of
ICA in analyzing the complex spatiotemporal dynamics of nonequilibrium
dissipative systems.
For the LCGL, ICA extracts a hole structure as a characteristic
feature, while for the NCGL, ICA extracts a gap structure with two
tails as a characteristic feature.
Although there exists no general analytic relationship, the ICA bases
obtained in this letter are similar to the wavelet
bases~\cite{Mallat,NCGL5} in the sense that the basis vectors are
localized both in space and in frequency, and cover the entire space
by spatial translation.
However, these ICA basis vectors do not form a hierarchy over scales
through dilatation, which is characteristic of the wavelet basis
vectors. The fractality of the amplitude pattern in the NCGL is
embodied in the singularity of each basis vector, as shown in
Fig.~\ref{fig_ncgl_power}.
In this letter, we dropped the phase information of the oscillators
for simplicity, but the phase also contains important information. In
order to capture the phase information, complex ICA of STC, which is
conceptually more difficult, is now under investigation.

We thank D. Tanaka, M. Yamada, Y. Iba, H. Suetani and A. Rossberg for
useful comments, K. Arai for proofreading the manuscript, and Yukawa
Institute of Kyoto University for providing computer facilities.

\end{document}